# Lattice eddy simulation of turbulent flows

**Jinglei Xu[1,3], Qi Li[1,2], Xianxu, Yuan[2*], Lin Bi[2*], Pengxin Liu[2], Jianqiang Chen[2]**

1Department of Energy and Power Engineering, Beihang University, Beijing 100191, China.

2State Key Laboratory of Aerodynamics, China Aerodynamics Research and Development Center, Mianyang 621000, China

3Hangzhou Innovation Institute Yuhang, Beihang University, Hangzhou 311100, China

Kolmogorov's (1941) theory of self-similarity implies the universality of small-scale eddies, and holds promise for a universal sub-grid scale model for large eddy simulation. The fact is the empirical coefficient of a typical sub-grid scale model varies from 0.1 to 0.2 in free turbulence and damps gradually to zero approaching the walls. This work has developed a **La**ttice **E**ddy **S**imulation method (LAES), in which the sole empirical coefficient is constant ($C_s$=0.08). LAES assumes the fluid properties are stored in the nodes of a typical CFD mesh, treats the nodes as lattices and makes analysis on one specific lattice, $i$. To be specific, LAES express the domain derivative on that lattice with the influence of nearby lattices. The lattices right next to $i$, which is named as $i+$, "collide" with $i$, imposing convective effects on $i$. The lattices right next to $i+$, which is named as $i++$, impose convective effects on $i+$ and indirectly influence $i$. The influence is actually turbulent diffusion. The derived governing equations of LAES look like the Navier-Stokes equations and reduce to filtered Naiver-Stokes equations with the Smagorinsky sub-grid scale model (Smagorinsky 1963) on meshes with isotropic cells. LAES yields accurate predictions of turbulent channel flows at $Re_\tau$=180, 395, and 590 on very coarse meshes and LAES with a constant $C_s$ perform as well as the dynamic LES model (Germano et al. 1991) does. Thus, this work has provided strong evidence for Kolmogorov's theory of self-similarity.

## 1. Introduction

Turbulence contains a large range of scale motions in both time and space. The enormous range of scales have to be resolved in the numerical simulation using the Navier-Stokes equations, and the computational cost is only affordable for flows with very low Reynolds numbers. The large eddies of the flow are dependent on the geometry while the smaller scales are more universal, according to the implication of Kolmogorov's (1941) theory of self-similarity. This feature allows one to explicitly solve for the large eddies in a calculation and implicitly account for the small eddies by using a sub-grid scale model (SGS model). The calculation, named as **L**arge **E**ddy **S**imulation (LES), allows the usage of a coarser mesh. However, Kolmogorov's (Kolmogorov 1941) theory of self-similarity seems to be invalid regarding the performance of sub-grid models. The Smagorinsky model (Smagorinksy 1963) is one typical sub-grid model in which the models "constant", $C_s$, can range from 0.1 to 0.2 in different flows. In addition, the constant needs to multiply an artificial damping function which varies from 0 to 1 in wall turbulence to reflect the wall damping effects. Germano et al. (1991) removed the empiricism by the implementation of a dynamic procedure which incorporates explicit filtering operations, ensemble averaging in homogeneous directions, and a somewhat ad hoc clipping to prevent an unstable (negative) eddy viscosity (Vreman 2004). Even though this dynamic model is one of the most accurate

†Email address for correspondence: yuanxianxu@cardc.cn
‡Email address for correspondence: bzbaby1010@163.com

SGS model, $C_s$ is also not constant and it varies in a mathematic manner rather than in a physical manner.

On the other hand, one may notice that LES is based on filtering, and the filter width lacks a rigorous definition. For example, some use $\Delta = \sqrt[3]{\Delta x \Delta y \Delta z}$ while others prefer $\Delta = \max(\Delta x, \Delta y, \Delta z)$. The choice of filter width matters especially when the grid anisotropy is large. In the following sections, this work will demonstrate that it is the ambiguous definition of filter width that leads $C_s$ to be non-constant.

## 2. One-dimensional analysis

The Navier-Stokes equations are the governing equations of fluid motions under continuum hypothesis. In the sense of continuum, the basic element of fluids is macromass which is with macro volume. In a typical CFD (Computational Fluid Dynamics) simulation, the mesh size is larger than the Kolmogorov scale. The Kolmogorov scale is larger than the macro volume, so the Navier-Stokes equations are not valid for such a coarse mesh. For simplicity, this work focuses on incompressible flows and performs analysis on a one-dimensional mesh as shown in Figure 1. Suppose the averaged fluid properties are stored on the grid point, and the local time derivative of velocity on the $i$ point is $\partial u / \partial t$. The fluid on the $i+1$ and $i-1$ points directly flows to the $i$ point, altering $u$ at the rate of $-u \partial u / \partial x$ which is exactly the convection term. The fluid on the $i+2$ and $i-2$ points directly flows to the $i+1$ and $i-1$ points, indicating the $i+2$ and $i-2$ points indirectly influence the $i$ point. The indirect influence is delivered by an "equivalent viscosity" with the dimension of [$u$][$l$], in which [$u$] represents the fluctuating velocity in the $x$ direction and [$l$] can be simply set to $\Delta x$. The equivalent viscosity, marked as $\nu_t$, satisfies dimensional analysis. The indirect influence is thus turbulence diffusion, and the turbulence diffusion rate on the velocity of the $i$ point is $\nu_t \partial^2 u / \partial x^2$. The $i+3$ and $i-3$ points also influence the $i$ point, but the influence are neglected considering third-order derivatives and $\Delta x^2$ are high order infinite small. One may assemble the above terms and the pressure force, arriving at the governing equation of velocity of the $i$ point,

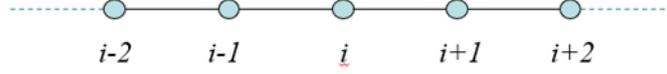

FIGURE 1. Illustration of one-dimensional mesh.

$$\frac{\partial u}{\partial t} = -u\frac{\partial u}{\partial x} + v_t \frac{\partial^2 u}{\partial x^2} - \frac{1}{\rho}\frac{\partial P}{\partial x} \qquad (1)$$

$u$ is velocity, $v_t$ is turbulence viscosity, $\rho$ is density and $P$ is pressure. Note that the $i$ point is an arbitrary point, so equation (1) is the governing equation for every point in the computational domain. In addition, due to the insufficient resolution of CFD grids, every grid point consists of many basic fluid elements. Among the basic fluid elements, the thermal motions of molecules cause inner viscous stress, *i.e.*, $v\partial^2 u/\partial x^2$. The $\partial^2 u/\partial x^2$ inside the grid point is unknown and is usually calculated by using the velocity of adjacent grid points, assuming the inside and outside second-order derivatives of velocity are identical. Thus, there are certainly discrepancies between CFD simulations and measurements. The viscosity, $v$, is with the dimension of $[u_0][l_0]$, in which $[u_0]$ is mean molecule velocity and $[l_0]$ is the free path of molecules. With the viscous stress taken into account, the governing equation becomes,

$$\frac{\partial u}{\partial t} = -u\frac{\partial u}{\partial x} + (v+v_t)\frac{\partial^2 u}{\partial x^2} - \frac{1}{\rho}\frac{\partial P}{\partial x} \qquad (2)$$

$$v_t = C\sqrt{k}\Delta x \qquad (3)$$

In equation (3), $k$ is the turbulent kinetic energy and at least an extra equation for $k$ must be solved. To avoid this, one could follow Smagorinsky's way (Smagorinsky ),

$$v_t = C_s^2 S\Delta x^2, \quad S = \sqrt{2S_{ij}S_{ij}}, \quad S_{ij} = \frac{1}{2}(\frac{\partial u_i}{\partial x_j} + \frac{\partial u_j}{\partial x_i}) \qquad (4)$$

## 3. Three-dimensional momentum equations

The analysis in section 2 is performed on the *i*th lattice and the nearby lattices. For three dimensional numerical simulations, the same analysis can be performed on the (*i, j, k*) lattice and the lattices around it, arriving at three dimensional momentum equations,

$$\frac{\partial u}{\partial t} + u\frac{\partial u}{\partial x} + v\frac{\partial u}{\partial y} + w\frac{\partial u}{\partial z} = -\frac{\partial p}{\partial x} + \nu\left(\frac{\partial^2 u}{\partial x^2} + \frac{\partial^2 u}{\partial y^2} + \frac{\partial^2 u}{\partial z^2}\right) + T_u \quad (5)$$

$$\frac{\partial v}{\partial t} + u\frac{\partial v}{\partial x} + v\frac{\partial v}{\partial y} + w\frac{\partial v}{\partial z} = -\frac{\partial p}{\partial y} + \nu\left(\frac{\partial^2 v}{\partial x^2} + \frac{\partial^2 v}{\partial y^2} + \frac{\partial^2 v}{\partial z^2}\right) + T_v \quad (6)$$

$$\frac{\partial w}{\partial t} + u\frac{\partial w}{\partial x} + v\frac{\partial w}{\partial y} + w\frac{\partial w}{\partial z} = -\frac{\partial p}{\partial z} + \nu\left(\frac{\partial^2 w}{\partial x^2} + \frac{\partial^2 w}{\partial y^2} + \frac{\partial^2 w}{\partial z^2}\right) + T_w \quad (7)$$

in which,

$$T_u = (C_s \Delta_x)^2 S\frac{\partial^2 u}{\partial x^2} + (C_s \Delta_y)^2 S\frac{\partial^2 u}{\partial y^2} + (C_s \Delta_z)^2 S\frac{\partial^2 u}{\partial z^2} \quad (8)$$

$$T_v = (C_s \Delta_x)^2 S\frac{\partial^2 v}{\partial x^2} + (C_s \Delta_y)^2 S\frac{\partial^2 v}{\partial y^2} + (C_s \Delta_z)^2 S\frac{\partial^2 v}{\partial z^2} \quad (9)$$

$$T_w = (C_s \Delta_x)^2 S\frac{\partial^2 w}{\partial x^2} + (C_s \Delta_y)^2 S\frac{\partial^2 w}{\partial y^2} + (C_s \Delta_z)^2 S\frac{\partial^2 w}{\partial z^2} \quad (10)$$

Equations (5)-(7) are identical to the traditional momentum equations of Smagorinsky LES on a mesh with isotropic cells. The Smagorinsky constant, $C_s$, is calibrated as 0.08. Note that, for a typical mesh with anisotropic cells, equations (8)-(10) yield an asymmetric subgrid stress matrix in the framework of LES. The authors think that the subgrid stress caused by insufficient mesh resolution are not inherent properties of fluids. The subgrid stress matrix depends on the distribution of cells, and anisotropic cells generates asymmetric matrix. One may keep in mind that equations (5)-(7) are not filtered momentum equations. The equations (5)-(7) are deduced via the analysis on lattices and the related simulation can be named as **La**ttice **E**ddy **S**imulation (LAES).

## 4. Simulation of turbulent channel flow

The incompressible turbulent channel flow is one classic canonical turbulent flow, and it is also one benchmark test case for the evaluation of numerical methods and SGS models. For the incompressible channel turbulent flow, one usually takes the wall friction velocity as the reference velocity and the half width of the channel as the reference length for the definition of the Reynolds number which reads, $\mathrm{Re}_\tau = u_\tau h / \nu$. The computational domain is a $2\pi h \times \pi h \times 2h$ (stream-wise length, span-wise width and height, respectively) box as shown in Figure 2. Periodic boundary conditions are used for the stream-wise and span-wise directions, and no-slip conditions are used on the walls.

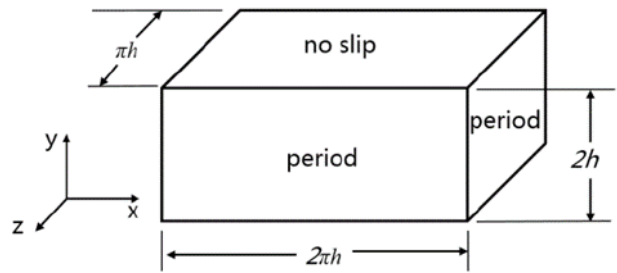

Figure 2. Computational domain and boundary conditions.

The CFD solver is LIEUT which was developed by researcher Xiaobing Deng in China Aerodynamics Research and Development Centre. In this turbulent channel flow solver, the projection method is used to solve the incompressible flow, and the space discretization adopts the hybrid method combining Fourier spectrum and high-order finite difference. The convection term is in Adams-Bashforth format with second-order precision, and the time derivative term and viscous term are discrete in Crank-Nicholson format. The time advancement step (non-dimensioned by unit grid length and unit velocity) is set to $5 \times 10^5$ for all the simulations. The lattice eddy simulations of the incompressible channel flow are performed at $\mathrm{Re}_\tau$=180, 395 and 590. For comparisons, large eddy simulations of the same flows are carried out using the classic Smagorinsky model (SM) and the dynamic Smagorinsky model (DSM). As shown in table 1, the simulations are performed on very coarse meshes which can provide extreme environments for the evaluation of simulation methods.

| Models \ Re numbers | LAES | Smagorinsky (SM) | dynamic Smagorinsky (DSM) |
|---|---|---|---|
| $Re_\tau = 180$ | 48 × 48 × 32 | 48 × 48 × 32 | 48 × 48 × 32 |
| $Re_\tau = 395$ | 48 × 56 × 40 | 48 × 56 × 40 | 48 × 56 × 40 |
| $Re_\tau = 590$ | 64 × 96 × 64 | 64 × 96 × 64 | 64 × 96 × 64 |

TABLE 1. Mesh resolutions for different Reynolds numbers (stream-wise grid number × wall-normal grid number × span-wise grid number)

The mean velocity profiles at different Reynolds numbers are shown in Figure 3-5. On such coarse meshes, the velocity profiles predicted by LAES agree very well with DNS data (Moser, 1999). The velocity profiles predicted by DSM are in reasonable agreement with DNS data. The mean velocity predicted by SM is too high. It is known that the total shear stress determines the velocity profile, and the total shear stress is composed of viscous shear stress, modeled subgrid shear stress and resolved shear stress. The resolved shear stress is only illustrated in Figure 6 to study different simulation methods, since LAES does not belong to the framework of subgrid model. The mesh for the $Re_\tau$=180 condition is with the largest grid anisotropy, and this condition is selected. The resolved shear stress predicted by DSM agree very well with DNS data, indicating that DSM mainly relies on sufficiently resolved shear stress to yield accurate velocity profiles. SM yield low resolved shear stress which corresponds to high velocity profiles. LAES yields high fluctuated stream-wise and wall-normal velocity correlation rather than "resolved shear stress". The root-mean-square(*rms*) velocities shown in Figures 7, 8, and 9 are more of physical significance for LAES. The stream-wise *rms* velocity predicted by LAES is generally lower than those predicted by SM and DSM. The wall-normal and span-wise rms velocities predicted by LAES is generally higher than those predicted by SM and DSM. In general, the velocity fluctuations released by DSM are in best agreement with DNS data, thanks to the dynamic procedure for calculating $C_s$. Nevertheless, the velocity fluctuation in the wall-normal direction released by LAES agree better with DNS data, indicating anisotropic filter width can improve the performance of SGS models. One may rely on DSM and LAES for the computation of turbulent noise. On the other hand, incorporating such dynamic procedure for calculating $C_s$ into LAES may further improve the performance of LAES.

Figure 10 and Figure 11 present the instantaneous stream-wise velocity contours for the $Re_\tau=590$ condition in the x–y cross section which is perpendicular to the wall-normal direction. LAES and DSM both yield turbulent streaks. It is evident that more small eddies are captured by LAES in both $y^+=10$ and $y^+=160$ sections, indicating that LAES is good at releasing turbulence dynamics and providing breathing images. This phenomenon seems to be against the observation that compared with DSM and SM, LAES yields less $u'_{rms}$. The authors infer that the generation of rich turbulence structures is due to that LAES yields more $v'_{rms}$ and $w'_{rms}$. Similar results can be obtained from the lower Reynolds number cases.

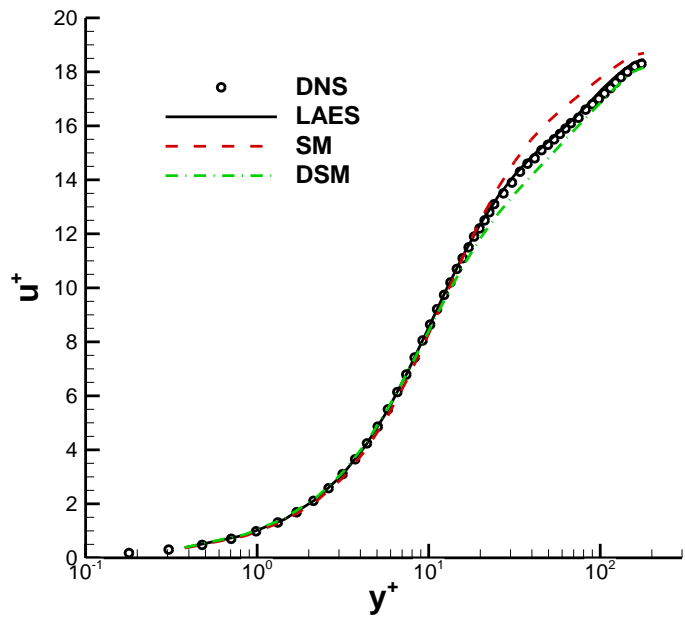

FIGURE 3. Mean velocity profile at $Re_\tau=180$.

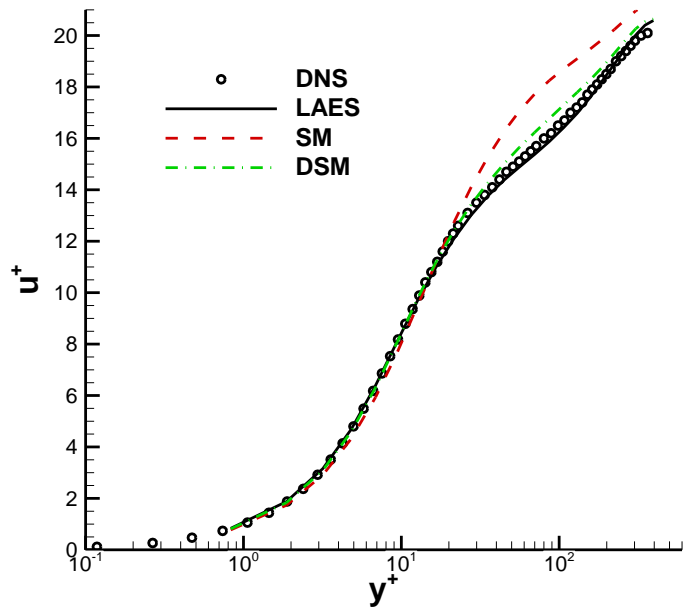

FIGURE 4. Mean velocity profile at $Re_\tau=395$.

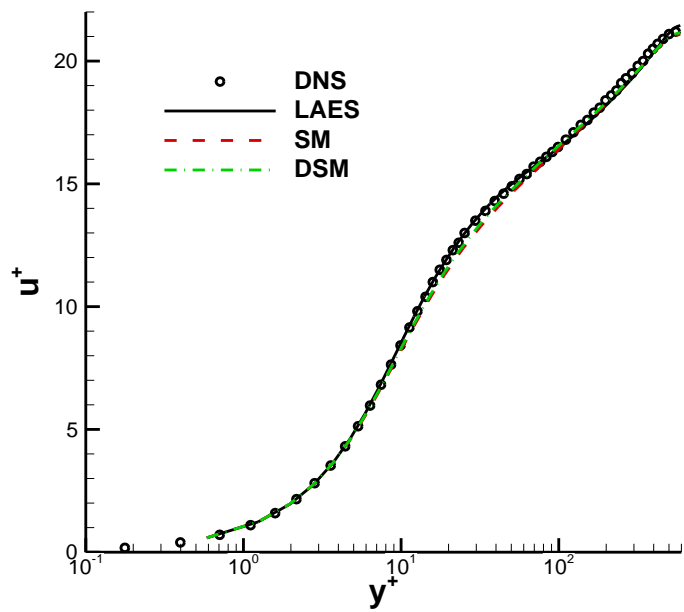

FIGURE 5. Mean velocity profile at $Re_\tau=590$.

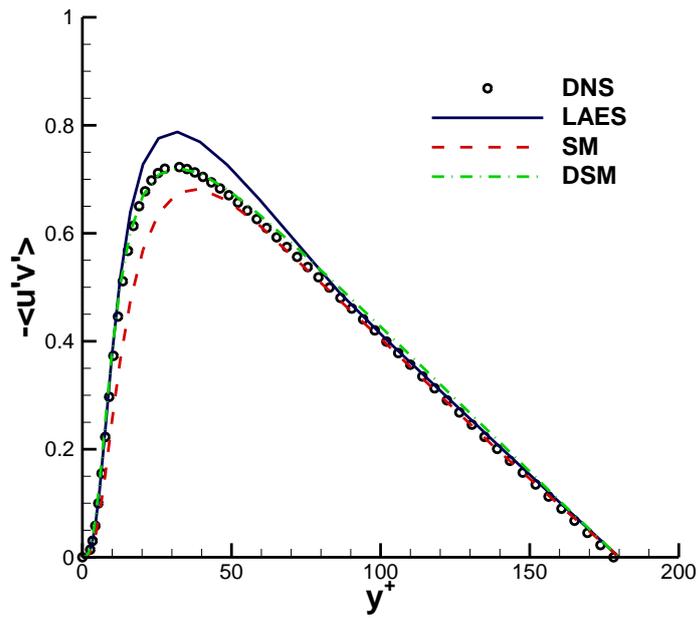

FIGURE 6. Reynolds shear stress profile at $Re_\tau=180$.

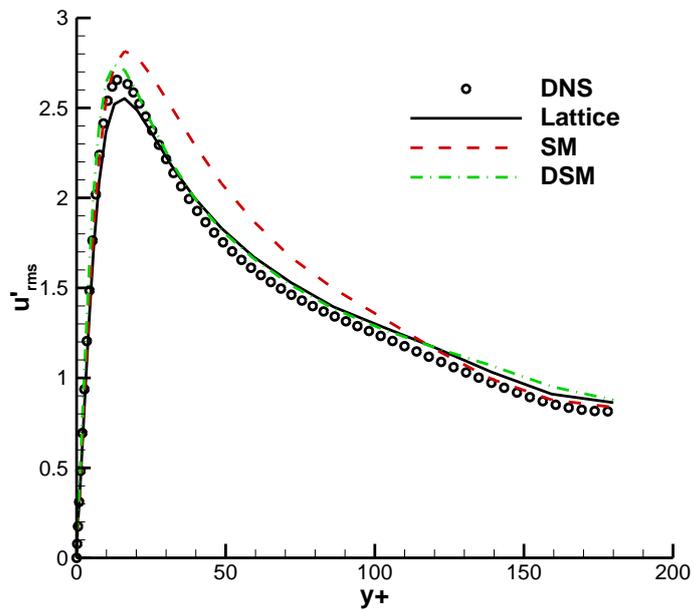

FIGURE 7. Magnitude of fluctuated stream-wise velocity profile at $Re_\tau=180$.

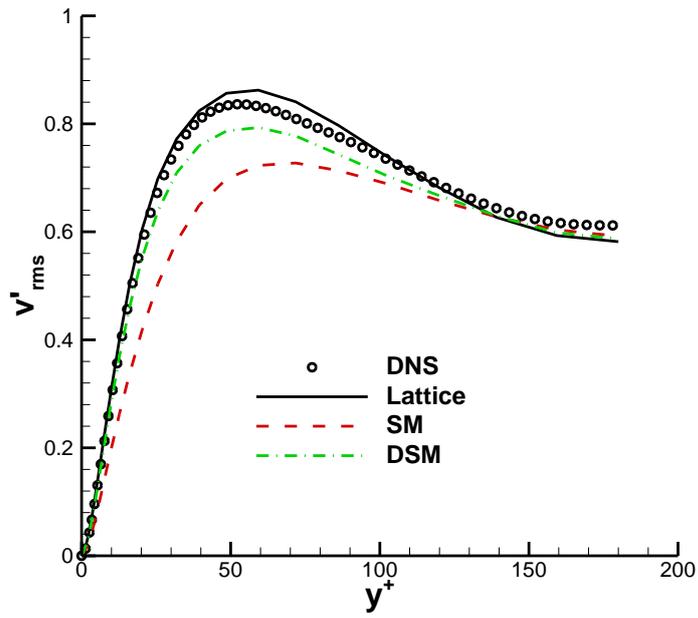

FIGURE 8. Magnitude of fluctuated wall-normal velocity profile at $Re_\tau=180$.

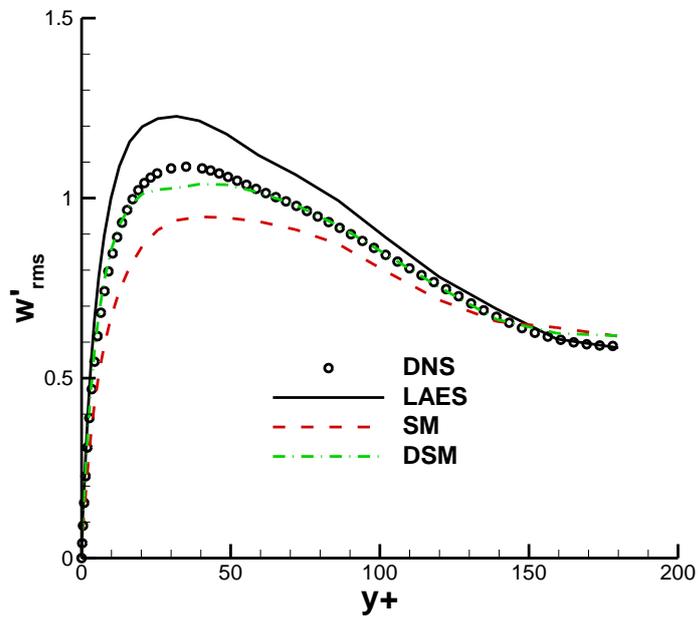

FIGURE 9. Magnitude of fluctuated span-wise velocity profile at $Re_\tau=180$.

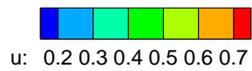

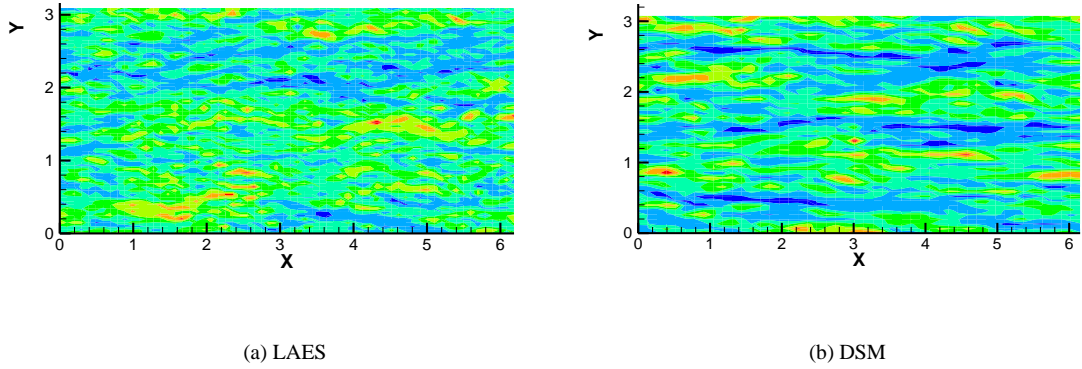

(a) LAES  (b) DSM

FIGURE 10. Contours of instantaneous streamwise velocity at $Re_\tau = 590$. The slices are extracted at $y^+=10$.

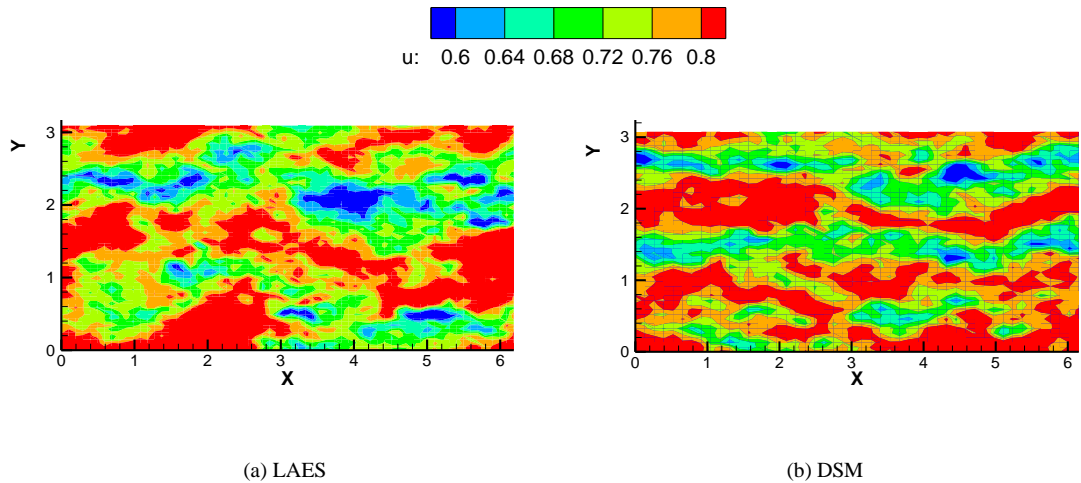

(a) LAES  (b) DSM

FIGURE 11. Contours of instantaneous streamwise velocity at $Re_\tau = 590$. The slices are extracted at $y^+=160$.

## 4. Conclusions

A novel LES-like turbulence simulation method has been presented. The governing equations are not filtered Naiver-Stokes equations but re-derived governing equations for fluid motions on one typical CFD mesh. The derivation focuses on one specific lattice and analyzes the influences of nearby lattices on the specific lattice. The derived governing equations are identical to that of the classic Smagorinsky LES on a mesh with isotropic cells, but generate asymmetric "subgrid stress" matrix on meshes with anisotropic cells. The asymmetric matrix, caused by anisotropic filter width, seems to be invalid but the application of the matrix on the turbulent channel flows justifies itself. Specifically, the mean velocity profiles predicted by LAES agree well with DNS data on very coarse meshes. The fluctuated wall-normal velocity predicted by LAES are in best agreement with DNS data, thanks to the wall-normal filter width being set to $\Delta y$. LAES also releases more turbulence structures in comparison with the predictions of DSM.

One may notice that LAES need not to use any *ad hoc* damping functions or any dynamic filtering procedure for wall turbulence. The model coefficient of LAES, also noted as $C_s$, is set to 0.08 in all the simulations. Thus, the only one model coefficient of LAES is truly a **constant**. And the authors are grateful that LAES has provided strong evidence for Kolmogorov's (Kolmogorov 1941) theory of self-similarity.